\documentclass[a4paper,11pt]{article}
\usepackage{jinstpub} 
\usepackage{lineno}
\usepackage{tabularx}
\usepackage{multicol}
\usepackage{caption}
\usepackage{siunitx}
\usepackage[position=top]{subfig}
\usepackage{floatrow} 
\notoc
\title{\boldmath The quality control programme for ITk strip tracker module assembly}

\author{A. Tishelman-Charny}
\affiliation{Brookhaven National Laboratory,\\
Upton, NY, USA}
\emailAdd{abraham.tishelman.charny@cern.ch}

\abstract{The assembly of the ATLAS Inner Tracker requires the construction of 19,000 silicon strip sensor detector modules in eight different geometries. Modules will be assembled and tested at 31 institutes on four continents from sensors, readout chips, and flexes. In order to adhere to the module specifications defined for sufficient tracking performance, a rigorous programme of quality control (QC) was established to cover components at every stage of assembly. This contribution presents an overview of the QC programme for ITk strip tracker modules, issues encountered during the pre-production phase (5\% of the production volume), and their solutions.}


\collaboration[c]{on behalf of the ATLAS ITk Collaboration}

\proceeding{Proceedings for TWEPP 2023 \\ Topical Workshop on Electronics for Particle Physics\\
  October 1-6 2023\\
  Geremeas, Sardinia, Italy}

\begin{document}
\maketitle
\flushbottom

\newpage

\section{Introduction}
\label{sec:introduction}

In order to ensure the production of 19,000 silicon strip sensor detector modules for the ATLAS Inner Tracker (ITk) robust enough to remain operational throughout the span of the High-Luminosity Large Hadron Collider (HL-LHC), a rigorous programme of Quality Control (QC) was established. 

The ITk strip detector will be composed of a barrel section with two module geometries, and two end-caps with six module geometries. Although the module geometries vary, their assembly procedures are extremely similar as their designs follow the same paradigm: ASICs are glued onto hybrid flexes, and hybrid flexes and powerboards are subsequently directly attached to silicon strip sensors with an adhesive. Wire-bonds are subsequently made between the strips and ASICs.

In order to ensure no issues are introduced during assembly, each step has a defined set of QC procedures to follow. 

This paper is structured as follows: section~\ref{sec:Module_structure} will describe the structure and various geometries of ITk strip modules. Section~\ref{sec:Assembly_and_QA_QC_steps} will describe module assembly and QC steps. Section~\ref{sec:Issues_encountered} will describe the issues encountered during QC and their solutions, and Section~\ref{sec:conclusions} will present the conclusions. 


\section{Module structure}
\label{sec:Module_structure}

\begin{figure}[h!]
    \centering
    \includegraphics[width=0.65\textwidth]{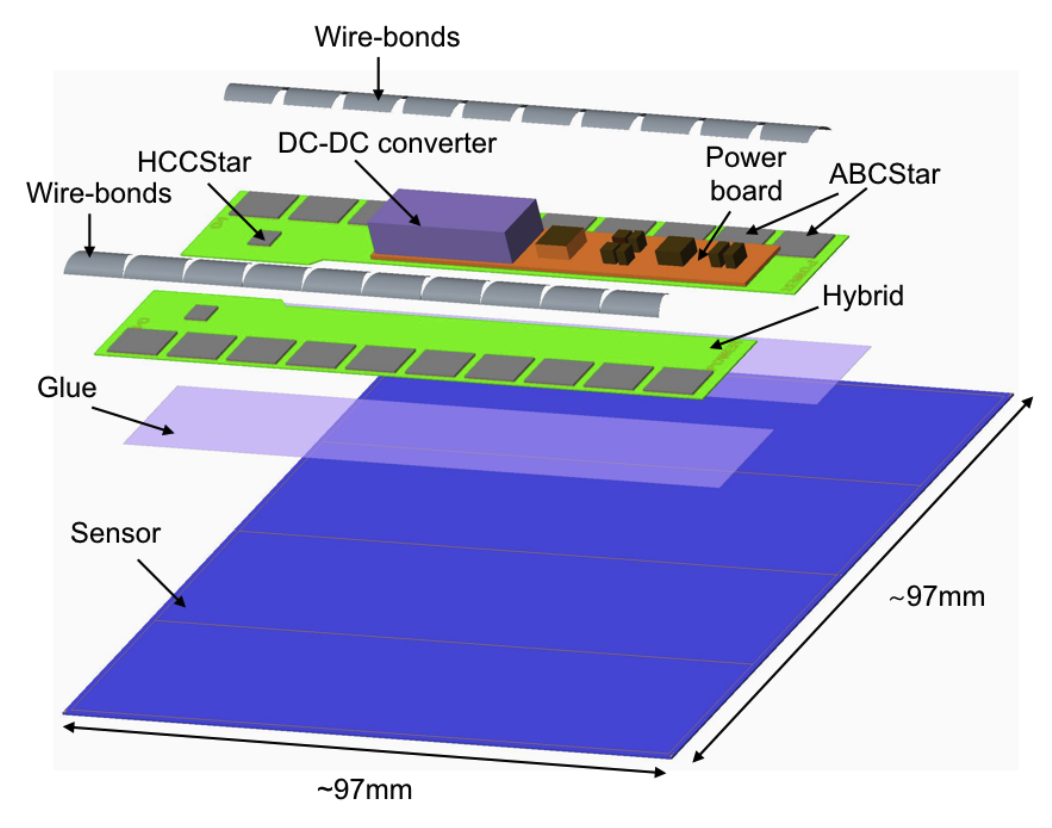}
    \caption{Diagram of a module \cite{TDR}}
    \label{fig:module_diagram}
\end{figure}

Modules are composed of silicon strip sensors serving as active detection material, with PCBs directly glued on top: hybrids hosting ASICs, directly wire-bonded to individual strips for readout, and a powerboard to power module components. As an example, a diagram of a module is shown in figure~\ref{fig:module_diagram}

\begin{figure}[htbp]
\centering
\includegraphics[width=0.55\textwidth]{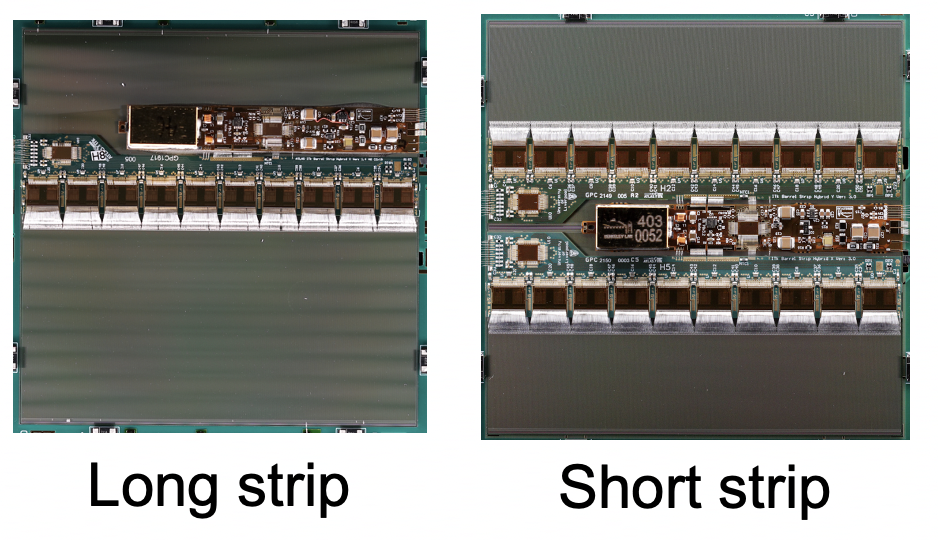}
\caption{\label{fig:Barrel_modules}Barrel modules. Note that short strip modules have two hybrids to provide readout of four sections of strips, while long strip modules only require one hybrid to read out two sections of strips.}
\end{figure}

\begin{figure}[htbp]
\centering
\includegraphics[width=0.85\textwidth]{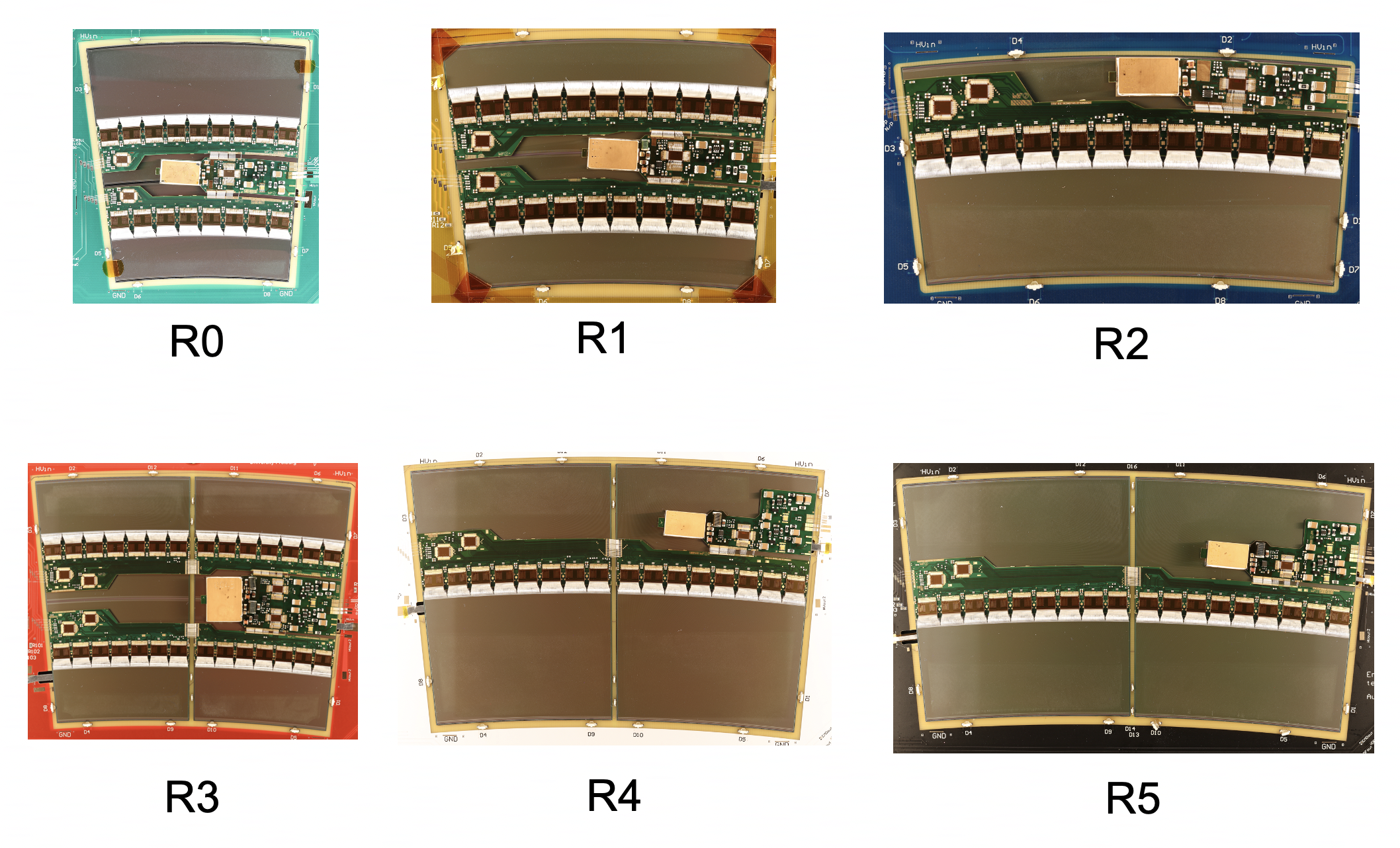}
\caption{\label{fig:Endcap_modules}End-cap modules (not to scale). Note that ring modules R3-R5 contain two sensors.}
\end{figure}

Images of barrel and end-cap modules are shown in figure~\ref{fig:Barrel_modules} and figure~\ref{fig:Endcap_modules}.

In the barrel, short strip modules will make up the two innermost layers, and long strip modules the outer two layers, as higher precision will be needed closer to the interaction point where a higher track occupancy is expected. In the end-cap disks, composed of petals, six module geometries are required in order to make up the petal shape.

\section{Assembly and QC steps}
\label{sec:Assembly_and_QA_QC_steps}

\noindent The assembly and QC procedure of ITk strips modules is identical for every step of barrel and endcap module assembly, except for the additional stitch bonding step required for ring modules R3-R5 composed of multiple sensors.

\begin{figure}[htbp]
\centering
\includegraphics[width=\textwidth]{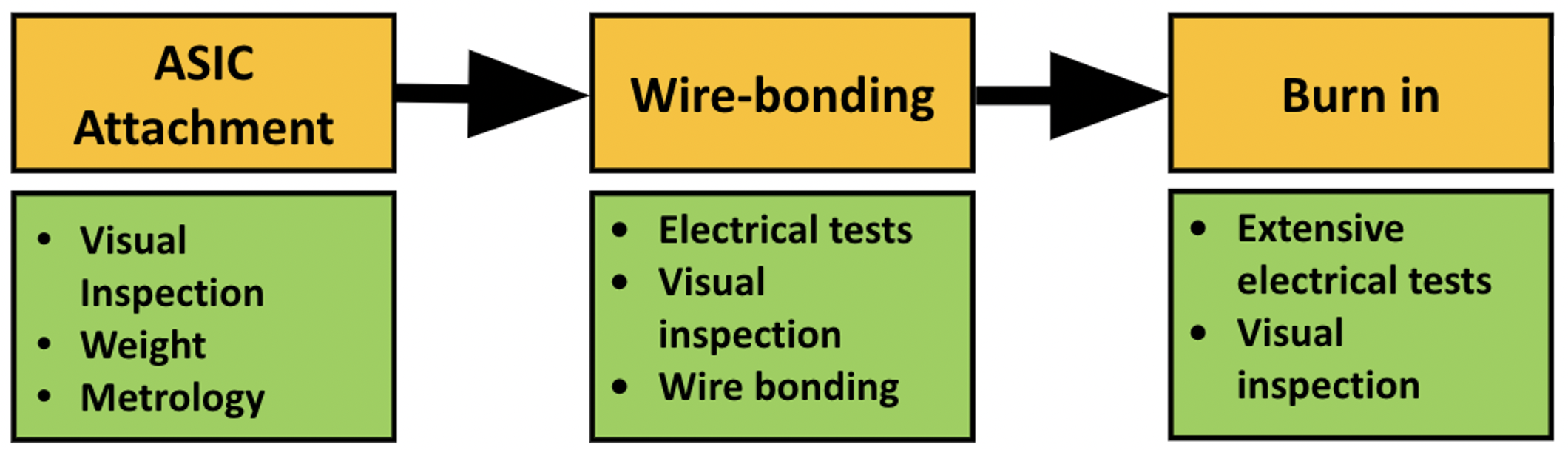}
\caption{\label{fig:Hybrid_Steps} Hybrid assembly and QC steps.}
\end{figure}

The hybrid assembly and QC procedure, shown in figure~\ref{fig:Hybrid_Steps}, corresponds to three stages before hybrids are glued onto modules. 

After each assembly step, a visual inspection is performed to spot obvious issues. These steps are defined as follows: first, ASICs are glued to hybrid flexes. Hybrids are then weighed in order to determine the amount of glue dispensed, as too much glue can lead to seepage, and too little glue can lead to insufficient support for wire-bonding. Metrology is also performed to ensure the ASICs are properly aligned for optimal wire-bonding. This involves ensuring ASIC xy-placement is within 200 microns of the nominal position, ASIC glue height is between 60-160 microns (compared to a nominal height of 120 microns), and the ASIC tilt angle is less than 0.025 to ensure vacuum quality is not affected during wire-bonding or module assembly. 

After wire-bonding the ASICs to the hybrid flex, an electrical test is performed to ensure proper electrical connection between the ASICs and hybrid using a software test suite \cite{ABC130}. During this test, an ASIC's readout channel (to later be wire-bonded to a single strip) is marked as bad if its gain is outside of the range 55-100  mV/fC, its noise outside 300-1700 ENC, or its gain is $>$ 3.5 $\sigma$ from the mean. A chip is considered bad if the standard deviation of its channels' gain (noise) values is $>$ 2.8 mV/fC (25 ENC). The standard deviation of input noise threshold of 25 ENC is currently being re-evaluated for end-cap modules, as different modules may require different thresholds due to their different geometries. Finally, a burn-in is performed corresponding to about 100 hours of electrical testing to catch hybrids which fail early. During burn-in, a channel fails if it is bad for more than 15\% of the electrical tests, and a hybrid fails if more than 1\% of its channels fail. After passing these QC stages, hybrids are glued onto sensors. 

\begin{figure}[htbp]
\centering
\includegraphics[width=\textwidth]{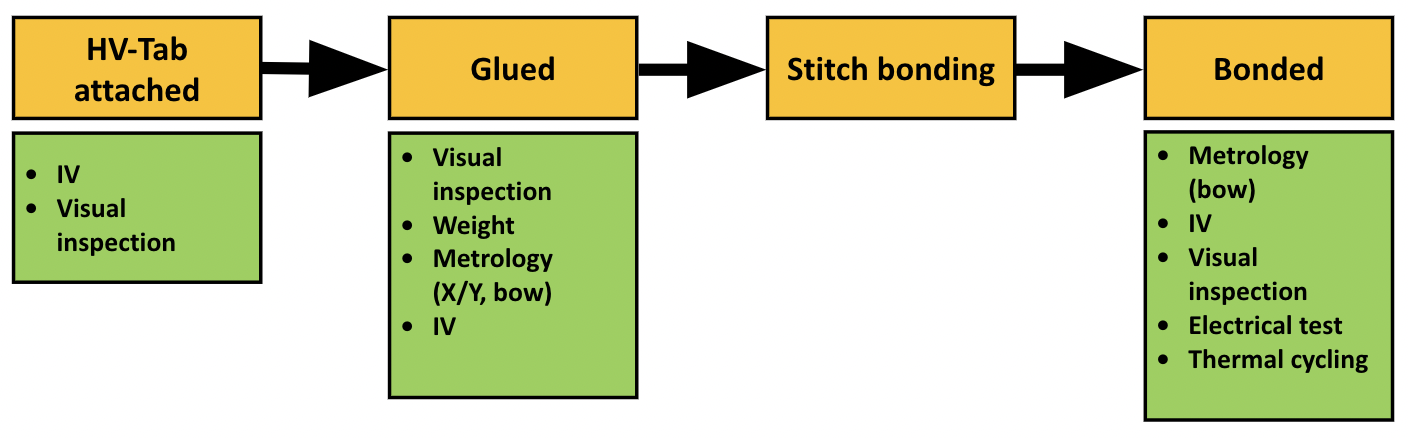}
\caption{\label{fig:Module_Steps} Module assembly and QC steps.}
\end{figure}

The module assembly and QC procedure, shown in figure~\ref{fig:Module_Steps}, begins with the HV-Tab attachment step: the HV-Tab is an aluminum strip attached to the metal backplane of the sensor in order to provide HV for sensor depletion. 

After each assembly step, the module is visually inspected in order to spot obvious issues. Step-specific assembly and QC steps are defined as follows: after attaching an HV-Tab to the sensor, an I-V curve is taken up to -700 V to catch HV-tabbed modules with early breakdown. A module fails the I-V step if it has a breakdown at $>$ -500 V. While initially modules can work with a high signal-to-noise ratio while biased at -350V, ensuring they are operational up to -500 V allows for the possibility to increase their bias voltage at the end of life. Hybrids and powerboards are then glued to the sensor, where a similar weighing and metrology procedure is performed as is performed on hybrids. This is to ensure there is not too much or too little glue, and to ensure the electronics are properly positioned for optimal wire-bonding between the ASICs and strips. After stitch bonding for multi-sensor ring modules, the final wire-bonding is performed between the ASICs and the strips. Metrology is performed to ensure sensors have not bowed, and multiple electrical tests are taken: first, an I-V curve up to -550V is taken to measure leakage current and check for breakdown. During the development of this test, it was found that at each voltage step an average of 100 current measurements needed to be used to extract stable I-V curves. An electrical test is then performed to characterize the module's noise. This test consists of a measurement of each ASICs timing offset and pedestal, the injection of three test charges with varied thresholds for noise measurement, and the injection of ten test charges for a more precise per-channel noise measurement. Further detail can be found in section 3.2.6 of \cite{ABC130}. A module fails this test if $>$ 2\% of its channels are marked as bad, or if 8 consecutive channels are bad. Finally, each module is thermal cycled ten times from $-35^\circ$C to $+20^\circ$C, and a noise measurement is taken after cycling. If the post-cycling noise measurement fails, the module fails thermal cycling. This step is performed to ensure the module is operational after the repeated temperature changes expected during operation at HL-LHC.

\section{Issues encountered}
\label{sec:Issues_encountered}

\noindent The purpose of the assembly QC procedure is to catch problems in assembled modules, be it systematic or statistical in nature. Identifying systematic issues in the assembly or testing procedure is a crucial benefit of QC. Three select issues encountered during the QC of pre-production modules are described below.

During the development of the thermal cycling module QC step, in which modules are thermal cycled 10 times to ensure they remain operational after the many temperature changes expected during the HL-LHC period, modules were found to exhibit large noise after going cold. This ``cold noise" prompted an extensive investigation, where it was discovered that capacitors on the barrel module powerboards were vibrating at the frequency of the DC-DC converter, about 2 MHz, leading to mechanical waves that traverse the sensor. These waves then appear to produce a voltage which induces noise - the mechanism is not yet fully understood \cite{ColdNoise}. The mitigation strategy used for long strip modules is a change of the glue between the sensors and PCBs from polaris to Loctite Eccobond F112, as explained in \cite{ColdNoise}. It should be noted that it is not yet understood why this particular glue choice reduces cold noise.

\begin{figure}[h!]
    \centering
    \includegraphics[width=\textwidth]{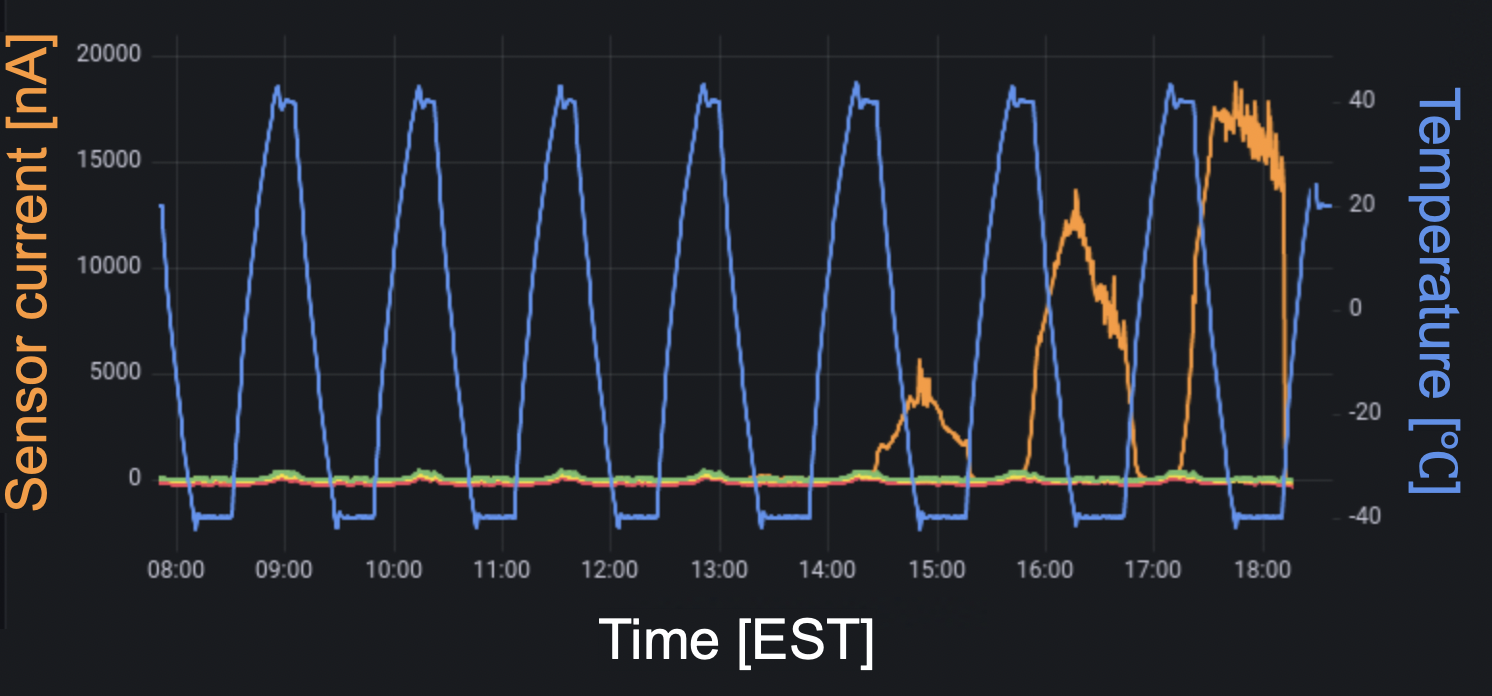}
    \caption{Sensor current vs. time during the first 7 cycles of a thermal cycle. Well-behaved sensor currents are drawn in green, yellow, and red.}
    \label{fig:TC_highCurrent}
\end{figure}

During further development and testing of the module thermal cycling sequence, some modules were found to display increasing sensor current with consecutive cycles. An example is shown in figure~\ref{fig:TC_highCurrent}.

During the 5th, 6th, and 7th thermal cycles of this long strip module's test, the sensor's current began increasing from about 5, to 10, and to 17 \si{\micro\ampere}. At the time, thermal cycling was performed with sensors biased at -350V throughout the sequence, including during temperature changes. Due to multiple instances of this behavior, and the fact that the detector is not expected to have HV on during temperature changes, the thermal cycling procedure was updated to only include HV powering during DAQ tests and not during temperature changes between cold and warm points. 

As another example, during the electrical testing of end-cap modules comprising two sensors, high noise was found in some segments. This high noise was traced back to the powerboard. This led to a re-design of all endcap module powerboards, as a solution to mitigate the noise \cite{EndcapNoise_PB}.

\section{Conclusions}
\label{sec:conclusions}

The pre-production phase of the ITk strip tracker module assembly, equivalent to about 5\% of the production volume, has been used to test the standard QC procedure. This experience has been crucial towards preparation for ITk strips production, as it has uncovered several issues and prompted appropriate solutions. In particular, the module electrical testing QC step revealed high noise in end-cap modules, and led to a redesign of the end-cap powerboards. Additionally, the module thermal cycling step revealed the existence of cold noise in some modules, which led to a change in glue choice between the sensor and PCBs. This step also revealed increasing leakage current during cycling of some modules, leading to the decision to bias sensors only during electrical tests and not while changing temperatures. 









\begin{thebibliography}{99}

\bibitem{TDR}
ATLAS collaboration, \emph{Technical Design Report for the ATLAS Inner Tracker Strip Detector}, \href{https://cds.cern.ch/record/2257755}{CERN-LHCC-2017-005} (2017)

\bibitem{ABC130}
L. Poley et al, \emph{The ABC130 barrel module prototyping
programme for the ATLAS strip tracker}, 2020 JINST 15 P09004 (2020)

\bibitem{ColdNoise}
I. Dyckes and M. Kurth on behalf of the ATLAS ITk collaboration, \emph{How the discovery of Cold Noise delayed the production of ATLAS ITk strip tracker modules by a year}, Topical Workshop on Electronics for Particle Physics 2023 (TWEPP 2023), Geremeas, Sardinia, Italy, 1 - 6 Oct 2023, these proceedings

\bibitem{EndcapNoise_PB}
D. Sperlich on behalf of the ATLAS ITk collaboration, \emph{Prototyping during pre-production}, Topical Workshop on Electronics for Particle Physics 2023 (TWEPP 2023), Geremeas, Sardinia, Italy, 1 - 6 Oct 2023, these proceedings

\end{thebibliography}
\end{document}